# Anomalous Creep as a Precursor to Failure in Granular Materials

Kasra Farain[1] and Daniel Bonn[1]

[1]Van der Waals–Zeeman Institute, Institute of Physics, University of Amsterdam; Science Park 904, 1098 XH Amsterdam, Netherlands.

**Abstract:** Granular materials, composed of discrete solid grains, can be modeled as simple mechanical systems. However, these materials can undergo spontaneous slow deformation, or creep, even under small forces and while in apparent mechanical equilibrium—a phenomenon central to understanding soil mechanics and the behavior of earthquake faults. We show that creep in granular materials originates from frictional dynamics at the contact points between grains. We reveal that the stability of these materials is governed by the interplay between creep and aging at these frictional contacts. Near the yield threshold, the frictional interactions result in anomalously accelerating creep, eventually leading to the delayed failure of the fragile packing. This behavior may serve as an early warning signal for catastrophic events like earthquakes and landslides.

**Main Text:** In most granular materials, the particles are typically large enough that thermally activated Brownian motion can be neglected. Consequently, in the absence of mechanical disturbances—which could cause a sudden rearrangement in the particles (*1*) or their gradual compaction (*2, 3*)—and when external forces are insufficient to induce large-scale flows, granular materials composed of *elastic solid* particles are generally expected to remain static over time. The elasticity of the particles is crucial. Granular systems in which particles can undergo viscoplastic deformation, vanish, or merge—such as foams, emulsions, and gel-based granular materials—behave differently. In these systems, local changes in particles can trigger long-range stress redistribution and particle rearrangements throughout the system (*4-8*), even without the presence of Brownian motion.

The physics of granular systems, consisting of elastic solid, non-Brownian grains, is particularly important due to its relevance to geohazards like earthquakes and landslides. The slow dynamics and stability of earthquake faults are heavily influenced by creep in a layer of granular material within the fault—called fault gouge—produced by comminution during sliding (*1, 9*). Similarly, soil, another non-Brownian granular system, undergoes gradual creep that shapes landscape evolution over geological timescales (*10, 11*) and contributes to landslides on steep hillslopes (*12*). Traditionally, soil creep has been attributed to physical and biological disturbances that churn the soil and increase its porosity (*3, 13, 14*). However, recent laboratory research has shown that even undisturbed sandpiles can exhibit creep, with rates and patterns closely resembling those observed on natural hillslopes (*3*).

In this study, we investigate the behavior of classical granular packings composed of hard Plexiglas grains (40 μm in diameter) subjected to constant stress levels below the material's yield stress in an isolated, noise-free environment. The yield stress marks the critical threshold at which a granular packing collapses, and the material begins to flow like a liquid (*15*). Using a simple rheometer setup, we conduct macroscopic-scale granular rheology experiments with nanoscale precision (see Methods). Despite the rigidity of the grains and the absence of external disturbances, significant creep deformation is observed. Our analysis reveals that this creep originates at the frictional contacts between grains in a configurationally stable granular packing. At these contacts, creep competes with frictional aging—the spontaneous increase in friction over time in stationary systems (*16-19*). This interplay allows us to identify the conditions under which the system transitions from typical logarithmic creep to anomalous creep behaviors, such as linear or accelerating creep, ultimately leading to delayed failure of the system.

We begin with a standard creep test, where a constant shear stress below the yield threshold is applied to the granular material, and its time-dependent deformation is measured. The results reveal a creep response that slows down in a logarithmic manner (Fig. 1). Generally, higher stress levels lead to increased deformation rates. However, the total deformation remains relatively small, typically less than 2% of the grain diameter over several hours.

From the perspective of classical mechanics, a system of rigid, non-thermal grains—regardless of their size or number—should quickly reach a stable mechanical equilibrium under sub-threshold forces. Such a system can be treated as a static assembly of rigid bodies governed by forces below the threshold of static friction. In this equilibrium state, the combined contact and frictional forces between grains counterbalance the external normal and shear loads. What, then, drives creep in an unperturbed, non-thermal granular system of rigid grains? Notably, this granular creep is at least four orders of magnitude greater than the bulk creep typically observed in Plexiglas (*20, 21*), ruling out the possibility that it arises from bulk deformation of the grains themselves.

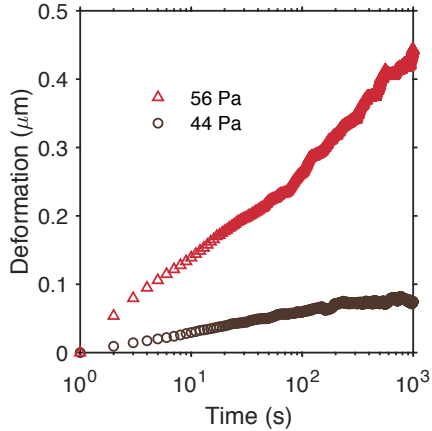

**Fig. 1. Creep deformation in granular materials.** Examples of creep deformation in randomly packed Plexiglas granules subjected to constant shear stresses of 44 Pa and 53 Pa.

The creep shown in Fig. 1 occurs in a granular material with an initially disordered grain arrangement along the shear plane. To investigate the mechanisms underlying creep behavior, we introduce an experimental protocol in which creep is observed in granular packings 'frozen' in their steady-state (ss) flow configuration. In these frozen flow configurations, the spatial arrangement of grains is no longer random, and transient start-up effects related to dilatancy, such as the well-known stress overshoot (*22*), are absent. To achieve this, we first apply a small, constant shear rate $\dot{\gamma}_0 = 0.0005$ s$^{-1}$, and allow the material to reach a ss-flow characterized by a ss-friction value of $\Sigma_{ss} = 46$ Pa (Fig. 2A). At time $t_0 = 0$, we then switch from imposing the shear rate to applying a constant shear stress of $\Sigma_1 = 44$ Pa, which is slightly below $\Sigma_{ss}$. This causes the flow to stop but the grain arrangement from the flow configuration remains intact.

The preservation of the grain flow configuration in the transitioning from the imposed strain rate $\dot{\gamma}_0$ to the imposed stress $\Sigma_1$ can be confirmed by reapplying $\dot{\gamma}_0$ after a brief interval (e.g., 1 s). In this case, the material immediately resumes ss-flow without any transient stress response, as it is already in a ss-flow configuration (*17*). This behavior can be understood by considering a free-body diagram of the grains under external forces. When transitioning from $\dot{\gamma}_0$ to $\Sigma_1$, such a free-body diagram would undergo negligible change. The shear stress on the material decreases only slightly, from $\Sigma_{ss} = 46$ Pa (the threshold stress required to sustain $\dot{\gamma}_0 = 0.0005$ s$^{-1}$) to $\Sigma_1 = 44$ Pa.

The frozen granular flow configuration under $\Sigma_1 = 44$ Pa exhibits logarithmic creep deformation similar to that observed in randomly packed configurations (Fig. 2A). This indicates that granular creep does not necessarily require the random spatial arrangement of grains or the presence of voids in the structure, as previously believed (*13*, *14*). During the creep, the material's strength also evolves. If, after the material has experienced creep in its arrested flow state for the duration $t_1 - t_0 = 400$ s, we reapply the shear rate $\dot{\gamma}_0$, a sharp frictional peak emerges (Fig. 2A).

We previously reported that the height of this friction peak increases logarithmically with the hold time $t_1 - t_0$ at a temperature-dependent rate (*17*). This same rate was also observed for stress relaxation in the material, highlighting a clear connection between the two phenomena. This can be understood as follows: The grains possess surface roughness, and under external load, the asperities at inter-grain contacts and the bulk of the grains deform (Fig. 2B) (*16*). These deformations store elastic energy in the material. During steady-state flow, the total elastic energy stored in the material remains constant (as it is a *steady* state). However, when flow pauses at $t_0 = 0$, the elastic energy stored in the frozen ss-flow configuration begins to dissipate via stress relaxation. After a waiting period, the frozen configuration contains less elastic energy than it did immediately after $t_0$, effectively sinking into an "elastic energy well" that deepens over time. To resume flow at $t_1$, an additional force is required to restore the lost elastic energy and lift the system out of this pinned, low-elastic-energy geometry. This extra effort manifests as the friction peak.

Hence, the observation of frictional aging denotes that the overall geometrical arrangement of grains and their inter-grain contacts remained unchanged during creep deformation. This conclusion is reinforced by the absence of any transient grain-scale stress response at $t_1$. Significant deviations from the ss-flow grain configuration during the creep interval from $t_0$ to $t_1$ would have inevitably induced stress responses over distances comparable to the grain size upon the reapplication of shear.

To further investigate the interplay between aging and creep, we once again froze the material in the ss-flow configuration (Fig. 2C). This time, however, at $t_1 = 400$ s, instead of reapplying the constant shear rate of $\dot{\gamma}_0$, we increased the shear stress to $\Sigma_2 = 53$ Pa, ensuring that $\Sigma_{peak}(400\text{ s}) > \Sigma_2 > \Sigma_{ss}$. Remarkably, despite the material being in a flow configuration without mechanical grain entanglements—such as those responsible for stress overshoot—it resists this shear stress greater than $\Sigma_{ss} = 46$ Pa due to the contact aging that occurred during $t_1 - t_0$; only slow logarithmic creep deformation is observed. At the end of the experiment, at $t_2 = 1200$ s, the shear rate $\dot{\gamma}_0$ is reintroduced to check for frictional aging. As before, this reveals a sharp frictional peak, with no detectable stress variations on the grain scale. Notably, this peak at $t_2$ (dark red) is larger than the one at $t_1$ (light red). These observations indicate that, throughout $t_0$ to $t_2$, the material maintained its ss-flow configuration with the same grain-to-grain contacts that continued to age. In the meantime, these contacts have necessarily experienced slight slipping to accommodate the total 400 nm of creep deformation. This deformation is distributed across approximately 50 grain layers in the granular bed, corresponding to an average of 8 nm of deformation per grain layer. For comparison, the diameter of the contact circle between grains under the ring's weight is estimated to be around 280 nm (see Fig. S2). Thus, while the grain contacts slip to facilitate the overall creep deformation, the slipping is minimal at each individual grain contact—insufficient to significantly alter the flow structure or disrupt the frictional aging at these contacts.

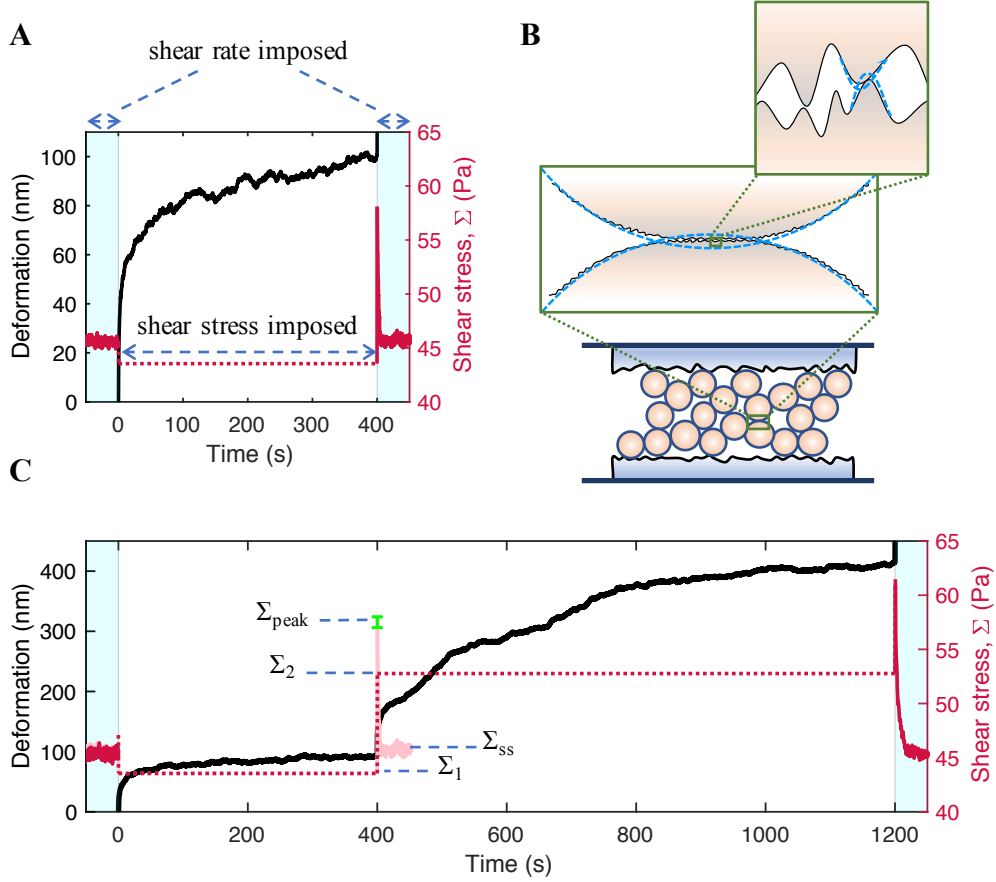

**Fig. 2. Creep deformation and frictional aging in arrested granular flows. (A)** Before $t_0 = 0$, the sample experiences steady-state flow under an imposed shear rate of $\dot{\gamma}_0 = 0.0005$ s$^{-1}$, with a shear stress response of $\Sigma_{ss} = 46$ Pa. At $t_0$, the system transitions to an imposed shear stress of $\Sigma_1 = 44$ Pa, slightly lower than $\Sigma_{ss}$, and the resulting strain response is recorded. Upon switching from imposing $\dot{\gamma}_0$ to imposing $\Sigma_1$, flow ceases, and only slow creep deformation is observed. At $t_1 = 400$ s, the shear rate $\dot{\gamma}_0$ is reapplied, revealing a narrow friction peak. **(B)** In a typical flow configuration where the material has frozen, surface asperities on regions of the grains in frictional contact undergo elastic deformation. The grains themselves also deform to conform to this flow configuration. Over time, stress relaxation within the deformed material reduces the total elastic energy associated with this specific grain geometry. The additional work needed to overcome the lost elastic energy and dislodge the system from this energy well manifests as a friction peak. Thus, the presence of a friction peak indicates a stable, pinned granular configuration during the interval $t_1 - t_0$. **(C)** The protocol in **A** is repeated until $t_1 = 400$ s. At $t_1$, the applied stress is increased to $\Sigma_2 = 53$ Pa, which remains below $\Sigma_{peak} = 58 \pm 0.5$ Pa, the friction peak resulting from 400 s of aging (green data point). The friction peak from **A** is also shown in lighter red. Since $\Sigma_2 < \Sigma_{peak}$, the granular system remains stable and only exhibits creep deformation. At $t_2 = 1200$ s, the stress is switched back to the shear rate $\dot{\gamma}_0$, revealing a larger friction peak than at $t_1 = 400$ s, indicating continued frictional aging from $t_0$ to $t_2$.

We repeat the experiment illustrated in Fig. 2C (without the final step at $t_2$ = 1200 s), incrementally increasing the shear stress, $\Sigma_2$, applied at $t_1$. The resulting creep deformations observed after $t_1$ are presented in Fig. 3. Although creep accelerates at higher stress levels, its overall behavior remains logarithmic until $\Sigma_2$ approaches $\Sigma_{peak}$ = 58 ± 0.5 Pa, the maximum strength imparted by frictional aging at grain contacts. This logarithmic trend reflects the system's eventual stabilization, as it means that the shear rate decreases inversely with time. However, as $\Sigma_2$ reaches $\Sigma_{peak}$, the creep transitions to a linear or even accelerating regime (Fig. 3, greenish-blue curve). This shift in creep behavior signals the system's approach to failure, even though deformation is still extremely slow—on the order of just a few nanometers per second.

At $\Sigma_2 \approx \Sigma_{peak}$, there exists an unstable balance between creep and frictional aging effects. Frictional aging (or stress relaxation) gradually strengthens the frozen ss-flow packing, while creep works to destabilize it. When creep surpasses frictional aging, the packing abruptly fails, undergoing rapid liquefaction. This is akin to an unstable equilibrium in a mechanical system, such as a ball precariously perched on the peak of a hill—any small disturbance causes it to roll downward. The liquefaction of the granular packing is marked by a sharp, uncontrolled spike in deformation rate, surging from a few nm/s to several mm/s. In our experiments, this delicate interplay between creep and aging can persist for a few seconds to several hundred seconds before failure occurs (see Figs. 3 and S3). Previous studies have visualized force-chain networks in granular packings subjected to external forces (*23, 24*). Sudden failure in fact occurs when the fragile force-chain network within the frozen ss-flow packing can no longer accommodate the ongoing creep deformation. Therefore, the delay time until failure may also depend on the properties of the ss-flow granular packing and its force-chain network.

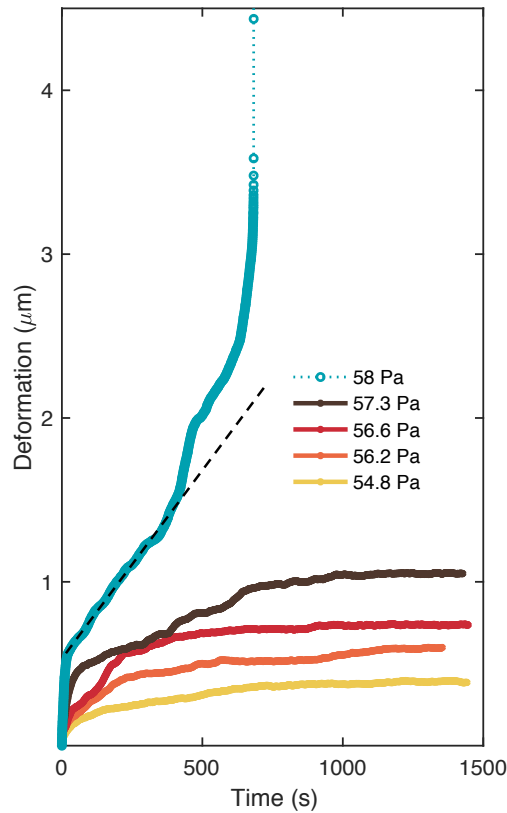

**Fig. 3. Abnormal creep and delayed failure in critically stressed granular packings.** Creep deformation as a function of time as $\Sigma_2$ (from Fig. 2C) increases toward the maximum strength of $\Sigma_{peak} = 58 \pm 0.5$ Pa, developed after 400 s of frictional aging in a granular packing arrested at the ss-flow condition. Initially, only the deformation rate increases with larger $\Sigma_2$, but eventually, the creep behavior shifts from logarithmic to a linear or accelerating trend as $\Sigma_2$ reaches $\Sigma_{peak}$ (greenish-blue curve). This accelerating creep culminates in spontaneous failure of the granular material (in this example, after 12 minutes), marked by a sharp increase in shear rate response from a few nanometers per second to millimeters per second and beyond.

The arrested flow scenario indeed arises naturally in fault gouge systems exhibiting stick-slip dynamics (*9, 25, 26*). During the 'stick' phase, the system remains in a pinned or arrested flow configuration, with stress gradually accumulating until it reaches a critical threshold, at which point the system slips. Our findings indicate that this transition is not instantaneous. Instead, as stress on the fault gouge increases, creep accelerates progressively, exhibiting a distinct qualitative change in behavior as it nears failure. These creep dynamics occur over extremely short length scales, making them challenging to detect. However, the slow movement of tectonic plates and gradual stress accumulation within Earth's layers may reveal such phenomena.

These findings also hold significant implications for understanding soil creep and modeling landscape evolution. Contrary to earlier assumptions (*3, 13, 14*), we have demonstrated that soil can undergo spontaneous creep without external disturbances or specific soil porosity conditions. This creep originates from minute movements at non-ideal, slowly adjusting frictional contacts between particles, accumulating across many layers of grains. Integrating this frictional mechanism into conventional hillslope creep models could enhance our understanding of sediment transport and provide valuable insights into how creeping soil may accelerate toward catastrophic landslides.

**Materials and Methods**

Our experimental setup involves a rheometer (MCR 302 Anton Paar) that rotates an aluminum ring (2.3 mm thick at the edge, with an outer diameter of 29.7 mm) floating on a ~2 mm thick bed of granular material (Fig. S1), as described elsewhere (*17*). This setup simulates dragging a 2.3-mm-wide sledge horizontally through sand using a rope, but without the leading or trailing edges that could cause ploughing effects (*27*).

The normal stress within the granular material remains perfectly constant throughout the experiments, determined solely by the weight of the ring (65 mN). This addresses a key limitation of typical rheometers, which are usually very precise in the shear direction but struggle to control the normal force. The MCR 302 Anton Paar rheometer is capable of rotating the macroscopic ring at remarkably low rotational speeds uniformly, reaching $6 \times 10^{-8}$ s$^{-1}$ easily, equivalent to edge speeds as low as 0.8 nm s$^{-1}$ (shear rate of $4 \times 10^{-7}$ s$^{-1}$ in our system) or applying torques with a precision of 10 nNm (corresponding to a shear stress of 3.7 mPa in our system). Additionally, it can also switch directly between the shear-rate- or shear-stress-controlled modes, which is essential for our experiments. The rheometer is placed on a vibration-isolated optical table in a quiet room. This setup is a laboratory surrogate for crustal faults, simulating the shearing of fault gouge between rock walls under the constant weight of overlying rock.

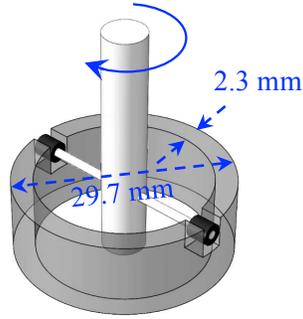

**Fig. S1.**
**Experimental setup.** Schematic representation of the rheology measuring system used in the study.

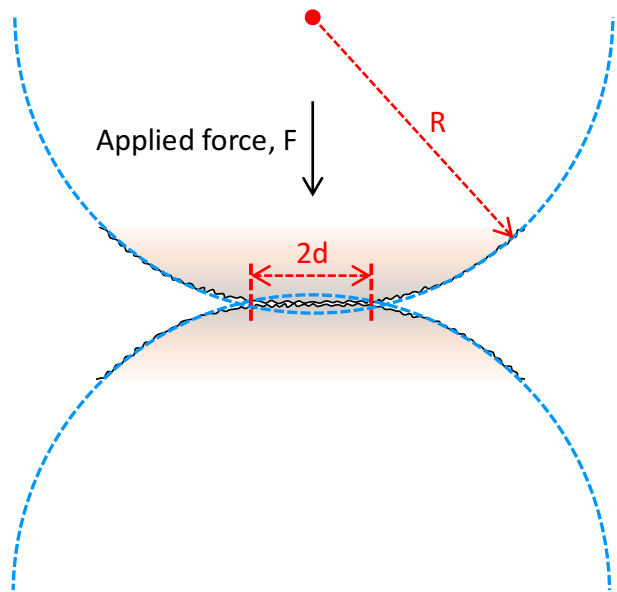

**Fig. S2.**
**Contact area between two grains.** When two elastic spheres are pressed together, their contact area is a circle. According to Hertz's theory for non-adhesive elastic contact, the radius $d$ of this contact circle is determined by the equation $d^3 = \frac{3FR}{4[E/(1-\nu^2)]}$, where $F$ is the force acting between the spheres, and $R$, $E$, and $\nu$ are the radius, elastic modulus, and Poisson's ratio of the spheres, respectively (*28*). The area beneath the ring shown in Fig. S1 is approximately 200 mm². Assuming a perfectly organized grain structure, this area could accommodate about 124,000 grains, each with a diameter of 40 μm. Given the ring's total weight of 65 mN, the average force supported by each grain is estimated to be 524 nN. Using the Hertz equation, the diameter of the contact circle is determined to be $2d$ = 280 nm. The elastic modulus ($E$) and Poisson's ratio ($\nu$) for Plexiglas grains are 3 GPa and 0.34, respectively.

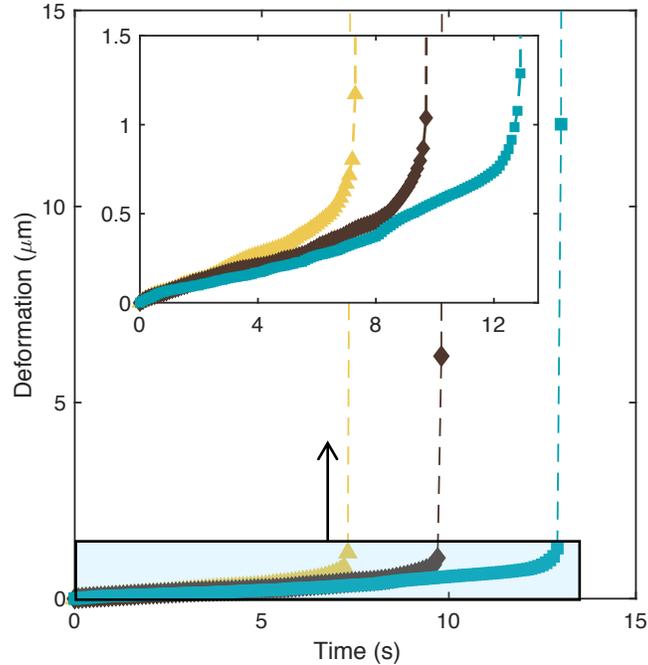

**Fig. S3.**
**Additional examples of over-logarithmic (linear or accelerating) creep behavior and delayed failure in granular packings under critical stress.** Creep deformation is plotted as a function of time for cases where $\Sigma_2$ (from Fig. 2C) reaches $\Sigma_{peak} = 58 \pm 0.5$ Pa, the maximum strength resulting from 400 s frictional aging at the frozen steady-state flow configuration

**Acknowledgments:** We thank B. Weber and J.A. Dijksman for insightful discussions.